\documentclass[aps,prx,reprint,superscriptaddress,floatfix]{revtex4-2}

\usepackage{graphicx}
\usepackage{dcolumn}
\usepackage{bm}
\usepackage{amsmath}
\usepackage{MnSymbol}%
\usepackage{wasysym}%
\usepackage{xspace}
\usepackage{gensymb}
\usepackage{textcomp}
\usepackage{xcolor}
\usepackage{mathtools}
\usepackage[mathlines]{lineno}
\modulolinenumbers[1]

\renewcommand{\eqref}[1]{Eq.~(\ref{#1})}
\newcommand{\para}{{\mkern3mu\vphantom{\perp}\vrule depth 0pt\mkern2mu\vrule depth 0pt\mkern3mu}}
\newcommand{\figref}[1]{{Fig.~\ref{#1}}}
\newcommand{\extfigref}[1]{{Fig.~\ref{#1}}}
\newcommand{\Figref}[1]{{Figure~\ref{#1}}}

\newcommand{\um}{\textmu{m}\xspace}

\newcommand{\rev}[1]{\textcolor{black}{#1}}
\newcommand{\SI}{\textit{Supplementary Information}\xspace}

\begin{document}
\title{Scaling transition of active turbulence from two to three dimensions}

\author{Da Wei}
\affiliation{Beijing National Laboratory for Condensed Matter Physics, Institute of Physics, Chinese Academy of Sciences, Beijing 100190, China}

\author{Yaochen Yang}
\affiliation{CAS Key Laboratory for Theoretical Physics, Institute of Theoretical Physics, Chinese Academy of Sciences, Beijing 100190, China}
\affiliation{School of Physical Sciences, University of Chinese Academy of Sciences, 19A Yuquan Road, Beijing 100049, China}

\author{Xuefeng Wei}
\affiliation{CAS Key Laboratory for Theoretical Physics, Institute of Theoretical Physics, Chinese Academy of Sciences, Beijing 100190, China}
\affiliation{School of Physical Sciences, University of Chinese Academy of Sciences, 19A Yuquan Road, Beijing 100049, China}
\affiliation{Wenzhou Institute, University of Chinese Academy of Sciences, Wenzhou, Zhejiang 325000, China}

\author{Ramin Golestanian}
\affiliation{Max Planck Institute for Dynamics and Self-Organization (MPIDS), D-37077 Göttingen, Germany}
\affiliation{Rudolf Peierls center for Theoretical Physics, University of Oxford, Oxford OX1 3PU, United Kingdom}

\author{Ming Li}
\affiliation{Beijing National Laboratory for Condensed Matter Physics, Institute of Physics, Chinese Academy of Sciences, Beijing 100190, China}
\affiliation{Songshan Lake Materials Laboratory, Dongguan, Guangdong 523808, China}

\author{Fanlong Meng}
\email{fanlong.meng@itp.ac.cn}
\affiliation{CAS Key Laboratory for Theoretical Physics, Institute of Theoretical Physics, Chinese Academy of Sciences, Beijing 100190, China}
\affiliation{School of Physical Sciences, University of Chinese Academy of Sciences, 19A Yuquan Road, Beijing 100049, China}
\affiliation{Wenzhou Institute, University of Chinese Academy of Sciences, Wenzhou, Zhejiang 325000, China}

\author{Yi Peng}
\email{pengy@iphy.ac.cn}
\affiliation{Beijing National Laboratory for Condensed Matter Physics, Institute of Physics, Chinese Academy of Sciences, Beijing 100190, China}
\affiliation{School of Physical Sciences, University of Chinese Academy of Sciences, 19A Yuquan Road, Beijing 100049, China}

\date{\today}
\begin{abstract}
Turbulent flows are observed in low-Reynolds active fluids. They are intrinsically different from the classical inertial turbulence \rev{and behave distinctively in two- and three-dimensions.} Understanding the behaviors of this new type of turbulence and their dependence on the system dimensionality is a fundamental challenge in non-equilibrium physics.
We experimentally measure flow structures and energy spectra of bacterial turbulence between two large parallel plates spaced by different heights $H$. The turbulence exhibits three regimes as H increases, resulting from the competition of bacterial length, vortex size and H. \rev{This is marked by two critical heights ($H_0$ and $H_1$) and a $H^{0.5}$ scaling law of vortex size in the large-$H$ limit.} Meanwhile, the spectra display distinct universal scaling laws in quasi-two-dimensional (2D) and three-dimensional (3D) regimes, independent of bacterial activity, length and $H$, whereas scaling exponents exhibit transitions in the crossover. To understand the scaling laws, we develop a hydrodynamic model using image systems to represent the effect of no-slip confining boundaries. This model predicts universal 1 and -4 scaling on large and small length scales, respectively, and -2 and -1 on intermediate length scales in 2D and 3D, respectively, which are consistent with the experimental results. Our study suggests a framework for investigating the effect of dimensionality on non-equilibrium self-organized systems. 
\end{abstract}
\maketitle



Turbulence is ubiquitous in nature, from galaxy formation, ocean current to human breathing. Besides the classical turbulence at high Reynolds number (\textit{Re}), turbulent flows are also discovered in active fluids at low \textit{Re}, such as bacterial suspensions~\cite{Dombrowski2004findBT,Peng2021,Liu21}, sperm swarms~\cite{Creppy2015spermTurbulence}, mixtures of microtubules and motor proteins~\cite{Martinez-Prat2019,Wu2017toroidFlow, Duclos2020}, epithelial cells~\cite{Blanch-Mercader2018,Lin2021}, and artificial motile colloids~\cite{Karani2019}. These active fluids are constituted of a large number of self-driven agents that inject kinetic energy individually into the systems. These agents generate active stress and drive the formation of turbulence. The internal-driven and self-organized nature distinguish this new class of turbulence from the classical ones. In this light, they are referred to as active turbulence~\cite{Alert2021review}.

Classical inertial turbulence exhibits universal scaling in kinetic energy spectra. In three dimensions (3D), turbulent flows are driven externally on macroscopic length scales and kinetic energy cascades towards small scales where it dissipates. On intermediate scales, vortices exhibit a scale-invariant structure, giving rise to a universal scaling exponent of -5/3~\cite{Kolmogorov1941}. When the systems are confined to be quasi-two-dimensional (2D), like large-scale motions in the atmosphere and oceans, dual cascade emerges with varied scaling because the vortex stretching is suppressed~\cite{Boffetta2012,Kellay2002}.

\begin{figure*}[htb!]
    \includegraphics[width=0.98\textwidth]{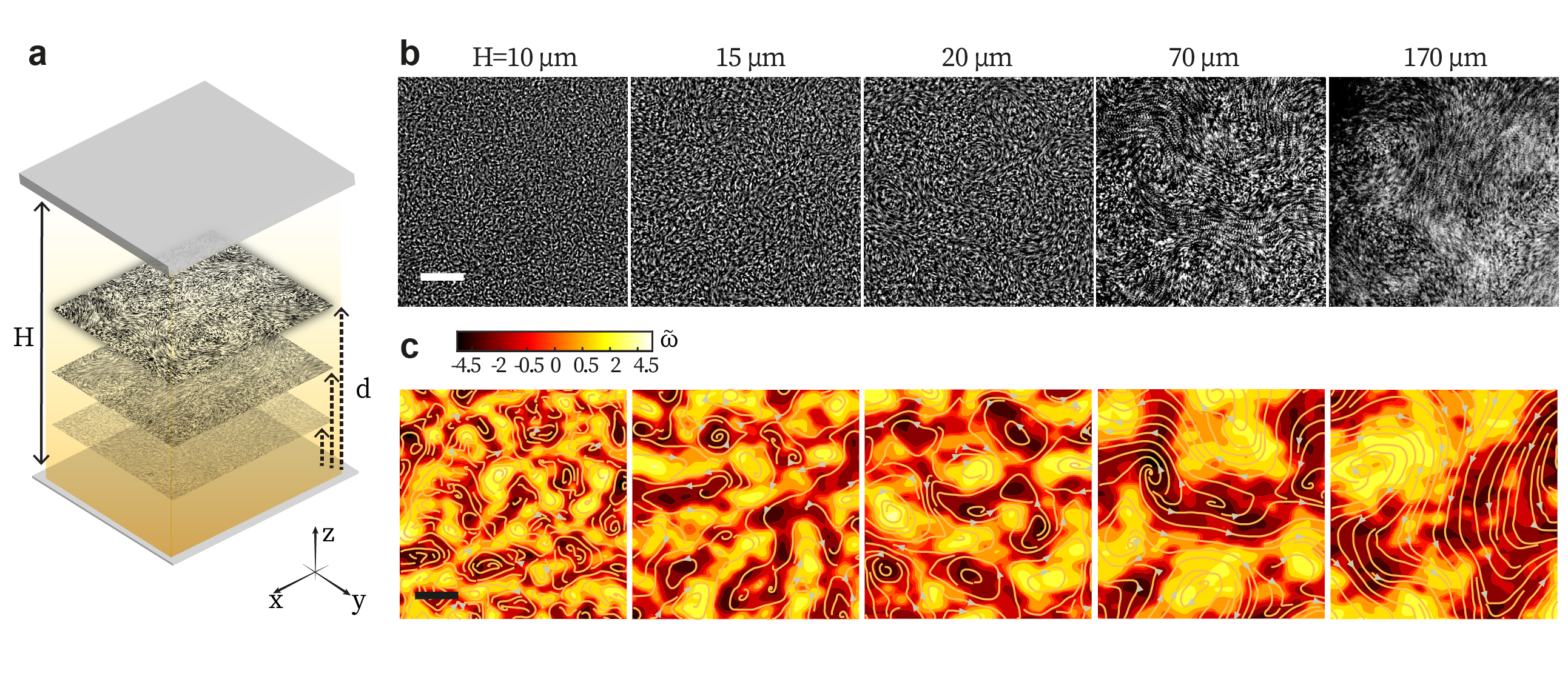}
    \vspace{-2mm}
    \caption{\textbf{Bacterial turbulence in samples of increasing confinement size} (a) Schematics of bacterial suspension confined between two parallel walls spaced by $H$. Bright field microscopy is performed at a distance $d$ from the bottom wall. (b-c) Typical flow patterns sampled at $d_c=H/2$ for different $H$. (b) The time-elapsed images. The elapsing time of each panel corresponds to $\sim5$\um/$\langle v\rangle$, where $\langle v\rangle$ is the mean flow field speed. (c) Corresponding vorticity field and streamlines.  $\Tilde{\omega}$ represents the vorticity scaled by its mean absolute value over the field ($\langle |\omega| \rangle$). Scale bars: 20~\um.}
    \label{fig:setup}
\end{figure*}

Compared with the better understood classical turbulence, whether universal scaling also exists in active turbulence is still under debate~\cite{Wensink2012,Bratanov2015,Martinez-Prat2019,Alert2020,Martinez-Prat2021prx}. 
Phenomenological models~\cite{Wensink2012,Dunkel2013,Heidenreich2016pre} extending Toner-Tu equations for polar flocking predict non-universal scaling laws in energy spectra~\cite{Bratanov2015}. On the contrary, theories considering the system as active liquid crystals suggest universal scaling~\cite{Giomi2015, Alert2020} for 2D active nematic turbulence, which is experimentally confirmed in the quasi-2D regime~\cite{Martinez-Prat2021prx}. 
Evidently, dimensionality and confinement play crucial roles in shaping the behaviors of active turbulence. So far, it has been shown that active turbulence bears different statistics in thin and thick samples~\cite{Wensink2012}; and confinement facilitates the coherent collective motion of active agents~\cite{Wioland2013,Lushi2014,Wioland2016,Wu2017toroidFlow,Martinez2020,Chandrakar2020prl,Varghese2020}.
Despite the consensus on the crucial role of dimensionality and confinement, distinguishing the 2D, 3D, and the crossover regimes remains to be a challenging task - let alone resolving the flow structures and scaling behaviors in these regimes.

Here we combine experiments and theories to resolve how active turbulence evolves from 2D to 3D and whether it exhibits universal statistical properties. We experimentally characterize turbulent flows in bacterial suspensions by measuring spatial velocity correlations, structure functions, and kinetic energy spectra with varying confinement heights. As a result, we uncover two critical heights that classify the turbulence into three regimes: the 2D, 3D, and an intermediate one. The two heights emerge from the competition between confinement and the cell size, and that between the confinement and the vortex size. Our results show that bacterial turbulence follows universal scaling laws in the 2D and 3D limits, independent of cell activity and length, and the system height. A hydrodynamic model is developed to understand the universal scaling in the 2D and 3D limits, as well as the transition between them, which is consistent with our experiments.

\section*{The structures of active turbulence}

We employ \textit{Escherichia coli} as our model system to study active turbulence (see Methods). Bacteria are suspended in a minimum motility buffer in which bacteria cannot grow further. To examine the effect of dimensionality, we inject bacterial suspensions into chambers confined in $z-$direction with different heights $H$ (2-400~\um, dimensions in $xy$-plane are 5-10~mm, \extfigref{fig:chamber}a). The chamber is sealed to avoid external flows and oxygen gradient. The bacteria concentration $\rho=3.2\times10^{10}$~cell/ml, which corresponds to bacterial volume fraction $\phi\approx6\%$. At this concentration, bacteria interact with their neighbours mainly by the hydrodynamic interactions~\cite{Peng2021}. We use video microscopy to image bacterial flow in $xy$-plane at a distance ($d$) from the bottom of the chamber (\figref{fig:setup}a). We perform particle image velocimetry (PIV) to extract the velocity field $\textbf{v}(\textbf{r})$ and compute the vorticity field $\omega(\textbf{r})$ (\figref{fig:setup}b-c, see also Methods).

\begin{figure}[htbp!]
    \includegraphics[width=0.50\textwidth]{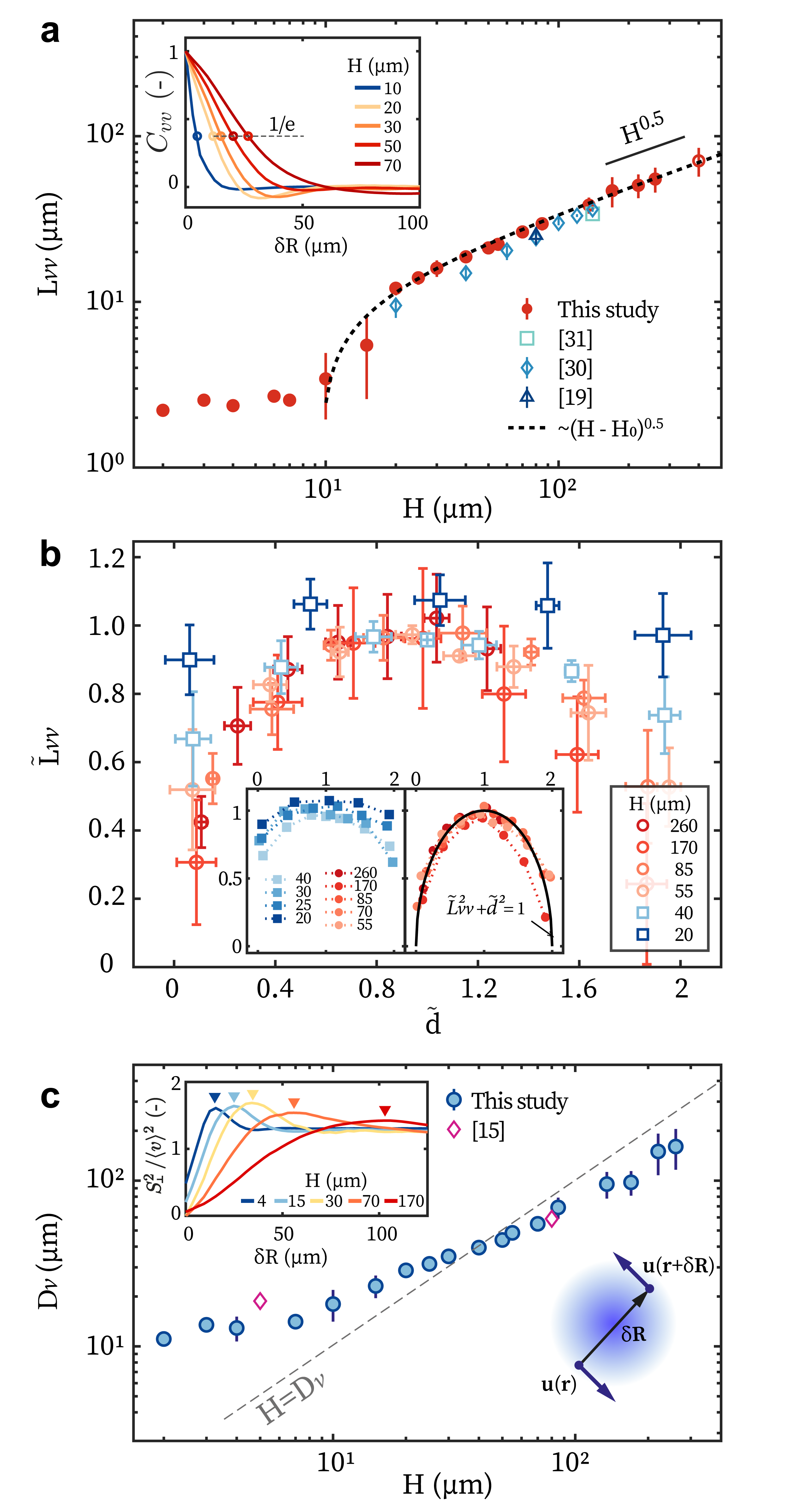}
    \vspace{-3mm}
    \caption{\textbf{The velocity structures of bacterial turbulence.} (a) $L_{vv}$ as a function of $H$. Data for $H=400$\um (hollow symbol) are experimentally estimated, see Methods. Inset: representative $C_{vv}$ as a function of $\delta R$; and $L_{vv}$ is defined as where $C_{vv}(\delta R)=1/e$. Dashed line: best fit with $L_{vv}=\sqrt{a(H-H_0)}$~\cite{Chandrakar2020prl}. (b) $\tilde{L}_{vv}=L_{vv}/\sqrt{a(H-H_0)}$ as a function of $\tilde{d}=2d/H$. Left inset: measured in samples with $H\leq40~$\um; right inset: with $H>40~$\um. (c) Effective vortex diameter $D_v$ as a function of $H$. Inset: typical $S^{2}_{\perp}$; and $D_v$ marks the peak position of $S^{2}_{\perp}$. Symbols and error bars in this figure represent mean$\pm$standard deviation. Data in (a) and (c) are taken at $d_c=H/2$. }
    \label{fig:Lvv}
\end{figure}

Bacterial turbulence exhibits distinctive flow patterns from 2D to 3D (\figref{fig:setup}, see also Supplementary Mov. 1-3). To characterize the evolving patterns, we first calculate the spatial velocity correlation functions, $C_{vv}(\delta R)=\langle\textbf{v}(\textbf{r})\cdot\langle\textbf{v}(\textbf{r}+\delta\textbf{R})\rangle / \langle \textbf{v}(\textbf{r})\cdot\textbf{v}(\textbf{r}) \rangle$, where $\delta R$ is the spatial distance and the angular brackets $\langle\cdot\rangle$ represents averaging over space and time. Representative $C_{vv}$ measured at the central height $d_c=H/2$ in samples of different $H$ are displayed in \figref{fig:Lvv}a inset.
The correlation length $L_{vv}$ is measured as the distance where $C_{vv}$ drops to $1/e$~\cite{Nishiguchi2015,Martinez-Prat2021prx}.
For a particular $H$, $L_{vv}$ is robust against variations in bacterial activity after the onset of turbulence~\cite{Sokolov2012,Heidenreich2016pre}, \rev{see also \figref{fig:activityIndependence}}. $L_{vv}$ increases negligibly when $H$ is below a critical height $H_0$, whereas it soars up as $L_{vv}=\sqrt{a(H-H_0)}$ above $H_0$ (\figref{fig:Lvv}a; $a=12.3~$\um, $H_0=9.6~$\um). Previous experimental measurements agree with our results~\cite{Dunkel2013,Guo2018,Liu2020}, which yet spans over a wider range of the system height. This allows us to resolve the critical length $H_0$ in the 2D samples, and to discover the scaling $L_{vv}\sim H^{0.5}$ in the 3D limit - which does not saturate up to $H=400~$\um (Methods). 
To understand this scaling, we conduct stability analysis on active polar systems confined by two no-slip walls (see \SI). We find that the confining boundaries make the system deviate from the long-wavelength instability predicted by the kinetic theory~\cite{Saintillan2008,Simha2002}. The wavelength corresponding to the fastest growth rate increases as $H^{0.5}$ in thick systems, which reflects the characteristic lengths of bacterial turbulence and underlies the observed trend in $L_{vv}(H)$. 

At a certain $H$, $L_{vv}$ varies symmetrically with $d$ (\figref{fig:setup}a) with respect to $d_c=H/2$ (\figref{fig:Lvv} and \extfigref{fig:LvvOverD}). We make $L_{vv}$ and $d$ dimensionless: $\tilde{L}_{vv}=L_{vv}/\sqrt{a(H-H_0)}$ and $\tilde{d}=2d/H$ (\figref{fig:Lvv}b). 
When $H$ is slightly above $H_0$, $\tilde{L}_{vv}$ depends weakly on $\tilde{d}$, e.g., when $H=20$~\um $\tilde{L}_{vv}$ varies less than $10\%$ over the entire $z$ domain (\figref{fig:Lvv}b inset).
As $H$ increases, the dependence of $\tilde{L}_{vv}$ on $\tilde{d}$ becomes stronger (\figref{fig:Lvv}b inset).
Interestingly, above a critical height $H_1\approx40~$\um, $\tilde{L}_{vv}$($\tilde{d}$) collapses to a master curve $\tilde{L}^2_{vv} + \tilde{d}^{\,2} = 1$, which is $H$-independent (\figref{fig:Lvv}b insets).
Such universal relation reveals scale-invariant turbulence structures for $H\geq H_1$.

We employ velocity structure functions to further resolve the structure of bacterial turbulence~\cite{Pope11,Kellay2002,Wensink2012}.
They average the two-point velocity difference $\delta \textbf{u}=\textbf{u}(\textbf{r}+\delta \textbf{R})-\textbf{u}(0)$ on orthogonal directions of $\delta \textbf{R}$ over the flow domain:
$S_{i}^n = \langle [\delta \textbf{u} \cdot \textbf{e}_i]^n \rangle\ \ (i=\perp,\para)$, 
where $n$ is the order and $\textbf{e}_{\perp,\para}$ are the unit vectors normal and parallel to $\delta \textbf{R}$ respectively. 
The maxima of $S_{\perp}^2$ correspond to an effective vortex size $D_v$ (\figref{fig:Lvv}c inset)~\cite{Pope11}. $D_v$ stays around 10~\um for $H\leq H_0$ and bears similar $\sim H^{0.5}$ scaling above $H_0$ (\figref{fig:Lvv}c). Importantly, $D_v(H)$ intersects with $H=D_v$ at $H_1$ (\figref{fig:Lvv}c), suggesting that $H_1$ emerges from the interplay between the vortex size and the confinement, e.g. due to the suppression of $z$-flows by the walls when $H<D_v$.
The near-wall hydrodynamics is long known to lead rod-shaped bacteria to swimming parallel to the wall~\cite{Berke2008}, and confinement smaller than vortex sizes rectify bacterial flow from turbulent to coherent (i.e. with uniform direction)~\cite{Wioland2013,Lushi2014}.
To identify this, we measure flow velocity in all three dimensions simultaneously near the focal plane with an enhanced particle tracking velocimetry (Methods and \extfigref{fig:PTV}). A strong suppression of $z$-flow is confirmed in thin samples, revealed by the large ratio of the $xy$-velocity to the velocity in $z$, $\langle u_{xy}/u_z\rangle$. The suppression is relaxed after $H \geq 40$\um$\approx H_1$ - as $\langle u_{xy}/u_z\rangle$ has reduced to the isotropic value $\sqrt{2}$~(\extfigref{fig:PTV}). Altogether, by analyzing velocity correlations and structures of active turbulence in real space, we classify the turbulent flows into three regimes: 2D ($H \leq H_0$), 2D-3D crossover ($H_0 < H \leq H_1$), and 3D ($H > H_1$).

\section*{Kinetic energy spectra and scalings}

\begin{figure*}[htbp!]
    \includegraphics[width=0.98\textwidth]{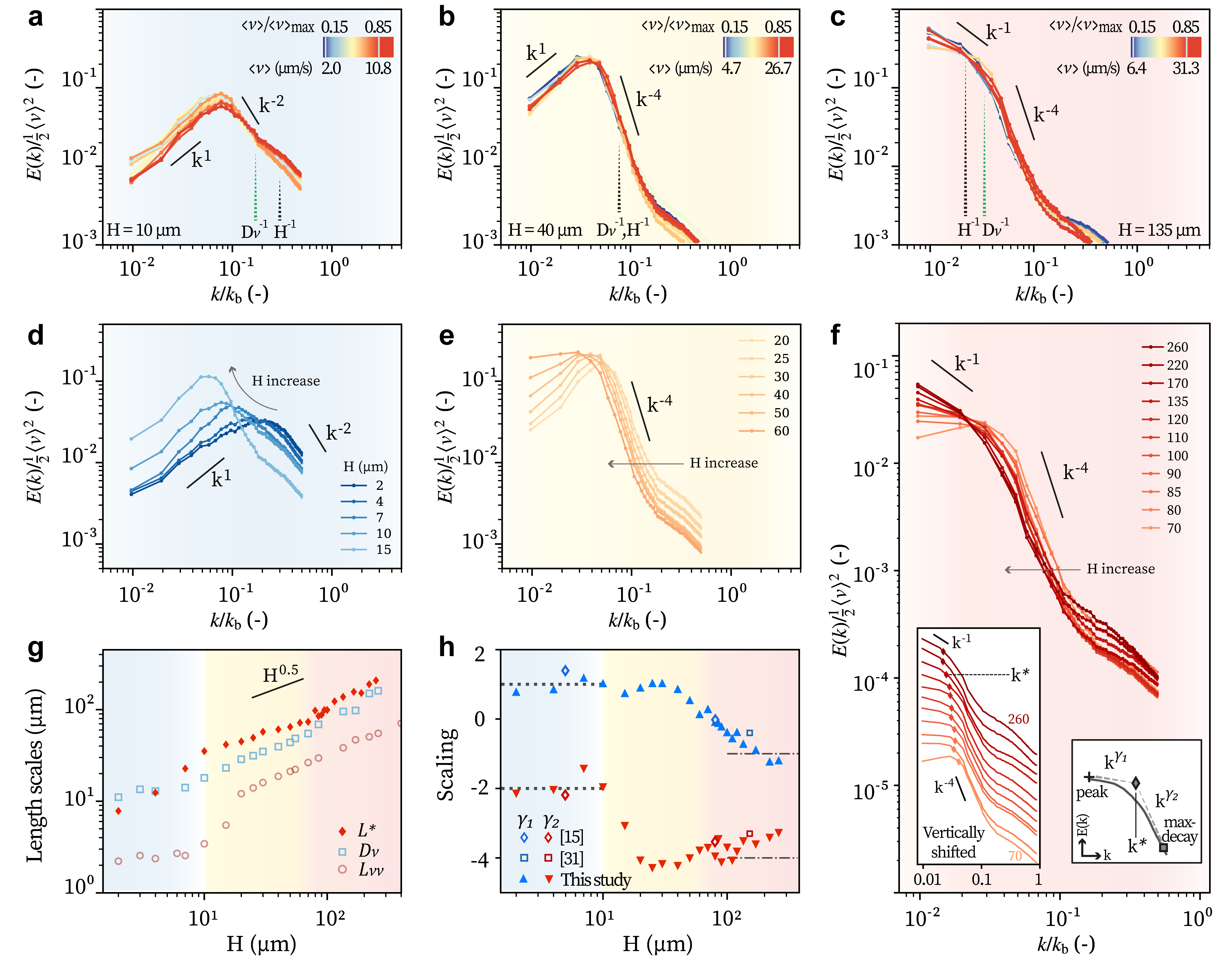}
    \vspace{-2mm}
    \caption{\textbf{Kinematic energy spectra of bacterial turbulence in different regimes.}
    Energy spectra for different activities ($\langle v \rangle$) at representative heights: $H\approx H_0$ (a), $H\approx H_1$ (b), and $H>H_1$ (c). $\langle v \rangle_{\rm max}$ is highest mean speed observed over multiple samples of the same height. The shown wavenumbers are scaled with $k_b=2\pi/L_{b0}$, where $L_{b0}=3$~\um is the bacterial length.
    Energy spectra evolving with $H$ for $H\lesssim H_0$ (d), $H_0\lesssim H\lesssim H_1$ (e), and $H>H_1$ (f). Every spectrum is averaged over the activity range $\langle v \rangle/ \langle v \rangle_{\rm max}>0.5$.
    (g) Characteristic lengths extracted. $L^{*}=2\pi/k^{*}$, with $k^{*}$ defined as shown in the schematics in f right inset. Left inset of (f): spectra vertically shifted for better visualization and with $k^{*}$ marked marked as diamonds. (h) Scaling exponents $\gamma_1$ and $\gamma_2$ as functions of $H$. Dotted and the dash-dotted lines: theoretical prediction of \eqref{eq:thinAndThickSample}. Blue, yellow, and red shading mark data obtained from 2D, 2D-3D crossover, and 3D regimes, respectively.}
    \label{fig:TKE}
\end{figure*}

We next study how kinetic energy of active turbulence is distributed on different length scales, by measuring the kinetic energy spectrum $E(k)$ defined as $\langle \textbf{u}^2 \rangle/2=\int E(k) dk$ in the wavenumber($k$)-space. $E(k)$ is reported for each height over multiple samples at varying bacterial activities. At a specific $H$, the dimensionless energy spectra $E(k)/\frac{1}{2}\langle u \rangle^2$ are independent of the activity (\figref{fig:TKE}a-c), showing the same scaling behaviors. Therefore, we report the dimensionless spectrum averaged over samples and activities for a certain $H$ (\figref {fig:TKE}d-f).  

In 2D ($H\leq H_0$), the energy spectrum is featured by two scaling regimes:$E(k)\sim k$ at low-$k$ (large length scales) and $\sim k^{-2}$ at high-$k$ (small length scales) (\figref{fig:TKE}a, d). Note that the rising tail at $k\gtrsim0.3k_b$ results from noise in velocimetry and microscopy in thicker samples. These spectra shift to lower $k$ (\figref{fig:TKE}d-f) as $H$ increases, which can be measured by the turning point $k^*$ between two regimes of different scaling exponents (\figref{fig:TKE}f right inset). The corresponding lengths $L^*=2\pi/k^*$ are displayed in \figref{fig:TKE}g. They both evolves as $H^{0.5}$ in the 3D limit, same as $L_{vv}$ and $D_v$. All these three quantities reflect the averaged length scales of unstable modes in bacterial turbulence at varying $H$, yet in different definitions or averaging methods.

The scaling laws transition from 2D to 3D in two steps: the high-$k$ scaling exponent $\gamma_2$ drops first from -2 to -4 at $H_0$, and then the low-$k$ exponent $\gamma_1$ changes from +1 to -1 when $H>H_1$ (\Figref{fig:TKE}h). The previous measurements on energy spectra are well in line with our results~\cite{Wensink2012, Liu2020}. Altogether, from the kinetic energy spectra we obtain the same statistical properties of active turbulence: the $H^{0.5}$ scaling of vortex size (\figref{fig:TKE}g and \figref{fig:Lvv}a) and the two critical heights $H_0$ and $H_1$ (\figref{fig:TKE}h and \figref{fig:Lvv}a-b).

\begin{figure*}[htpb!]
    \includegraphics[width=1\textwidth]{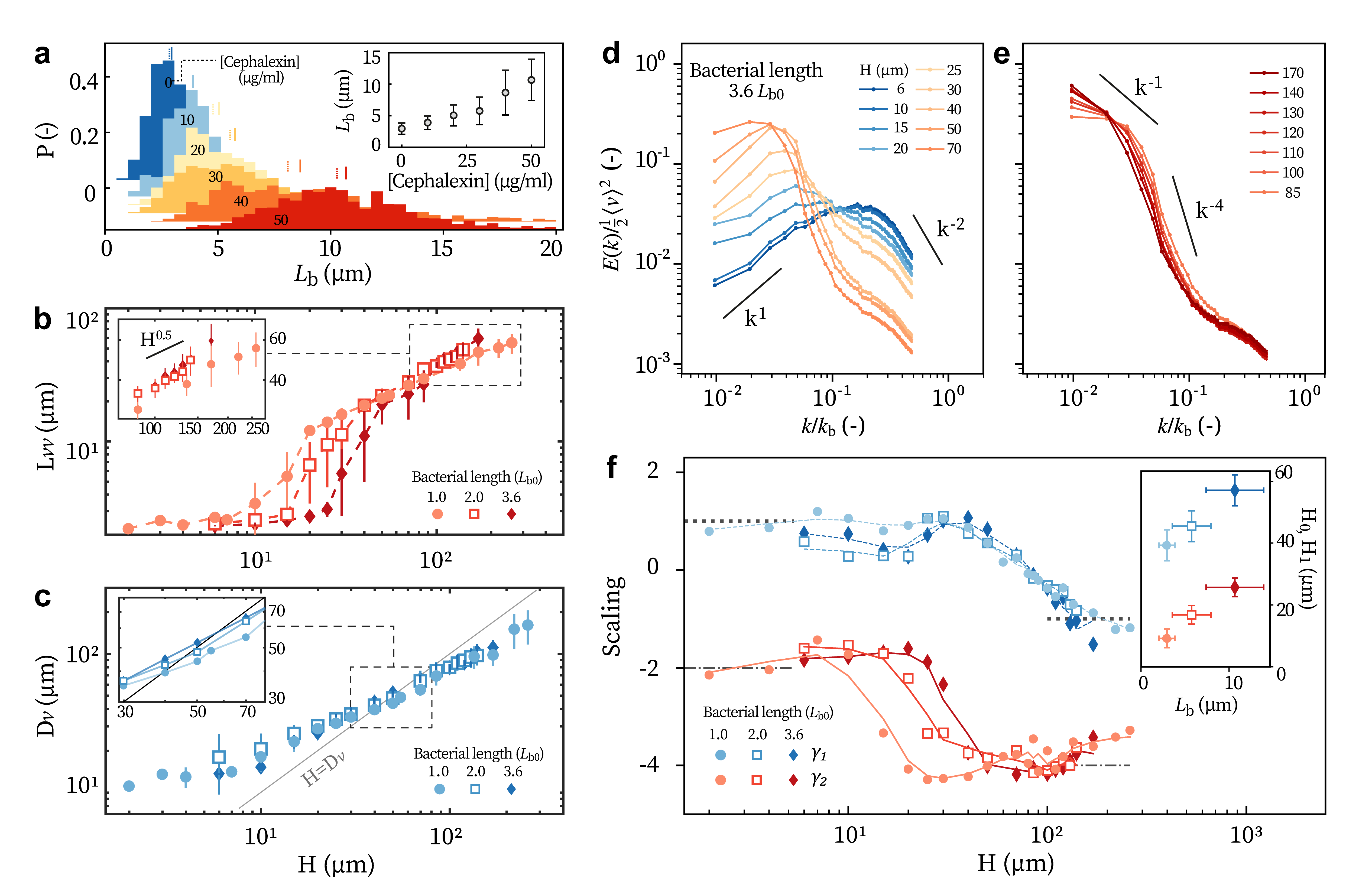}
    \vspace{-2mm}
    \caption{\textbf{Effect of bacterial length on the turbulence} (a) The distributions of cell lengths under different concentrations of cephalexin [Cephalexin]. Dotted and solid short lines respectively mark the median and mean values. Inset: mean value of $L_b$ for different [Cephalexin]. Error bars represent standard deviations. (b) $L_{vv}$ as a function of $H$ for different $L_{b}$. Inset: zoom in of the marked region. (c) $D_v$ as a function of $H$ for different $L_{b}$. (d-e) Energy spectra obtained with the most elongated cells ($3.6L_{b0}$). The blue-yellow transition corresponds to $H_0$. (f) Spectral scaling $\gamma_1$ and $\gamma_2$ obtained with elongated bacteria. Black dotted and dash-dotted lines: model predictions. Inset: $H_0$ and $H_1$ for different $L_b$. Error bars show standard deviations of $L_b$ ($x$-axis) and uncertainties in measuring $H$ ($y$-axis). All measurements are obtained at $d_c=H/2$ with $\phi=0.06$.}
    \label{fig:cellLength}
\end{figure*}

\section*{The effect of cell length on active turbulence}

Besides the system height and the vortex size, the cell length $L_b$ is another fundamental length scale of the active turbulence system. To clarify how $L_b$ affects the turbulent behaviors, we vary bacterial length by adding cephalexin, an antibiotics that prohibits cell division~\cite{Nishiguchi2015, Li2019bacteriaNematics} (Methods). $L_b$ increases monotonically with the cephalexin concentration [Cephalexin], and meanwhile no obvious changes are found in the size distributions (\figref{fig:cellLength}a and \extfigref{fig:cellLengthHistogram}) and the motility of a bacterium ~\cite{Nishiguchi2015,Li2019bacteriaNematics}. We experiment with bacteria whose mean length $\langle L_b \rangle\approx2.0L_{b0}$ (5.8~\um, [Cephalexin]=30~\textmu{g/ml}) and $3.6L_{b0}$ (10.7~\um, [Cephalexin]=$50~$\textmu{g/ml}) at the same bacterial volume fraction (6\%). 

\Figref{fig:cellLength} shows the correlation length, vortex size, and energy spectra for different bacterial lengths. Varying $L_b$ does not alter the $H^{0.5}$ scaling of $D_v$ or $L_{vv}$ (\figref{fig:cellLength}b-c). Energy spectra measured with elongated cells show the same scaling exponents in low-$k$ and high-$k$ regimes, with $(\gamma_1, \gamma_2)=(+1, -2)$ in 2D and (-1, -4) in 3D, respectively (\figref{fig:cellLength}d-e). This supports the universality of the scaling laws in bacterial turbulence. However, the critical heights increase for longer bacteria. $H_0$ follows an empirical relation $H_0= 2.1 L_b + 2.3~$\um and $H_1$ increases similarly with $L_b$ (\figref{fig:cellLength}b, f). Additionally, $H_0$ and $H_1$ measured in the representation of velocity structures ($L_{vv}$ and $D_v$) and in the representation of energy spectra display excellent equivalence (\extfigref{fig:H0H1equivalence}).

\section*{Hydrodynamic theory and universal scaling}
\begin{figure}[htb!]
    \includegraphics[width=0.48\textwidth]{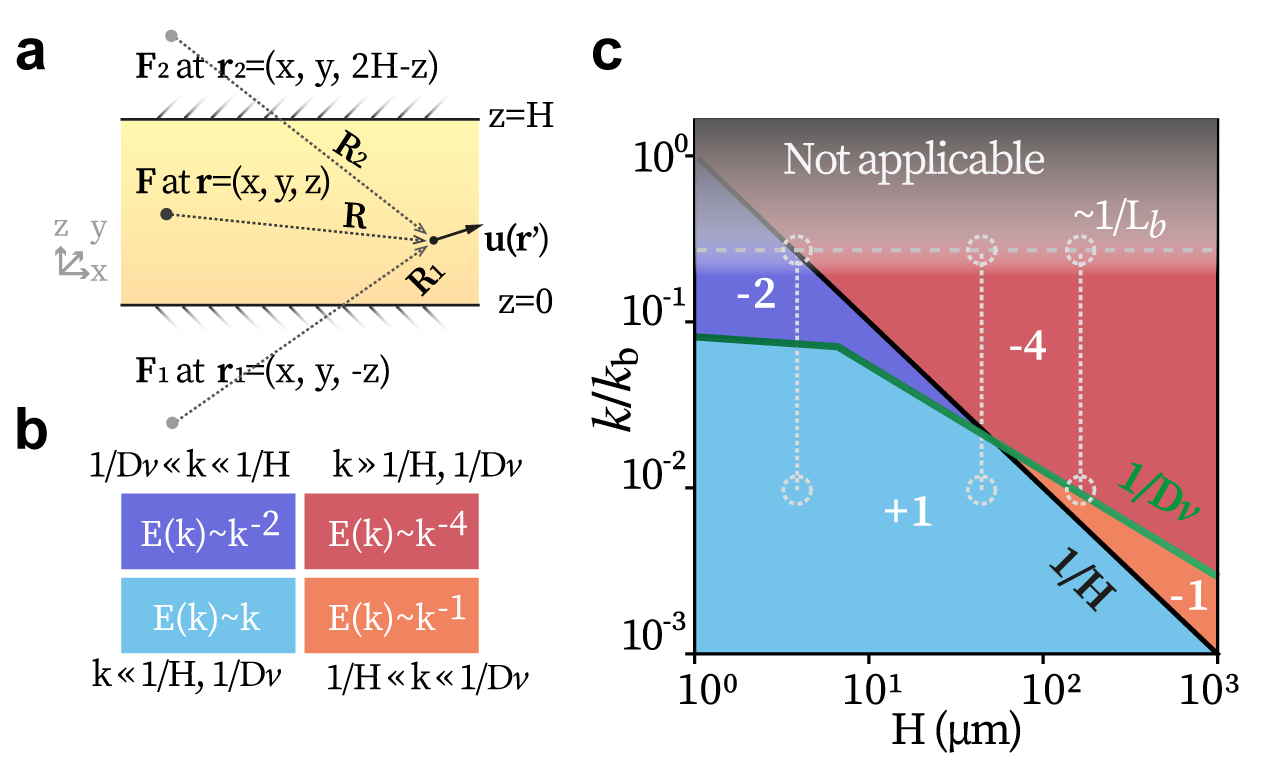}
    \caption{\textbf{Scaling regimes predicted by the hydrodynamic model.} (a) Scheme of the hydrodynamic model, where boundary effects are approximated by two images. (b) Asymptotic scaling behaviors results from the competition among $k$, $H^{-1}$, and $D_v^{-1}$. (c) Predicted scaling regimes for the bacterial turbulence. The green lines represent the $D_v - H$ relation. Each regime is marked by its scaling exponent.
    Vertical dashed lines: (from left to right) measured ranges corresponding to \figref{fig:TKE}a-c respectively.} \label{fig:theory}
\vspace{-3mm}
\end{figure}

To understand the scaling behaviors of 2D and 3D active turbulence, we develop a hydrodynamic model for active fluids confined by two no-slip boundaries. The model considers the active fluid as a continuum force field, and represents the effects of the upper and the lower walls by two image systems. A Green function with the first two image reflections is employed to approximate the images for a stokeslet with two no-slip boundaries~\cite{Liron1976,Meng2021,Ishida2022}, see \figref{fig:theory}a and \SI for more details. The energy spectrum $E(k)$ written in the form of the tensor is:
\begin{equation}
E(k)\sim k\langle |\hat{\textbf{u}}(\textbf{k})|^2\rangle 
    = k\bar{G}_{\alpha\beta}\bar{G}_{\alpha\gamma}\langle F_{\beta}F^*_{\gamma}\rangle,
\end{equation}
where $\bar{G}_{\alpha\beta}$ ($\alpha,\beta,\gamma=x,y$) is the Green function in Fourier representation, and $F$ is the active force density. The asymptotic behavior of $\bar{G}_{\alpha\beta}$ is:  
\begin{equation}\label{eq:Gab_asympt}
\bar{G}_{\alpha\beta} \sim \begin{dcases} 
(\delta_{\alpha\beta}-\frac{k_\alpha k_\beta}{2k^2})\cdot\frac{-H}{k}, & kH\ll1;\\
(\delta_{\alpha\beta}-\frac{k_\alpha k_\beta}{k^2})\cdot\frac{2}{k^2}, & kH\gg1.
\end{dcases}
\end{equation}
Hence, correspondingly, $E(k)$ behaves as:
\begin{equation}\label{eq:thinAndThickSample}
    E(k)\sim \begin{dcases}
    \frac{3H^2\eta^2k}{4}\langle\omega^2(\textbf{k})\rangle + 
    \frac{H^2}{4k}\langle F^2(\textbf{k})\rangle, & kH\ll1;\\
    \frac{\eta^2}{k}\langle\omega^2(\textbf{k})\rangle, & kH\gg1.
\end{dcases}
\end{equation}
Here $\eta$ is the shear viscosity and $\omega(\textbf{k})$ the vorticity field in Fourier space. Furthermore, the scaling behavior of $\langle\omega^2(\textbf{k})\rangle$ is governed by the competition between $k$ and the wavenumber corresponding to the vortex size ($D_v^{-1}$). Under the assumptions that the active vortices are uncorrelated and exponentially distributed in size (\extfigref{fig:ExpDistribution}), $\langle\omega^2(\textbf{k})\rangle$ scales respectively as $k^{0}$ and $k^{-3}$ for $kD_v\ll1$ and $kD_v\gg1$~\cite{Giomi2015}. Altogether, in the limit $F \ll \eta k \omega$,
the interplay among $k$, $H^{-1}$, and $D_v^{-1}$, give rise to four different scaling regimes, see \figref{fig:theory}b. We implement the empirical relation between $D_v$ and $H$ from the experiments (\figref{fig:Lvv}c) and display the predicted scaling regimes for our system in \figref{fig:theory}c.

The $k^1$- and $k^{-4}$-scaling laws are revealed to be universal, and they govern the kinetic energy distribution on the large and small scales, respectively. On large scales ($k\ll H^{-1},D_v^{-1}$), the growth of long-wavelength modes are suppressed by the presence of two no-slip boundaries, and such size selection underpins the $k^1$-scaling. On the small scales ($k\gg H^{-1},D_v^{-1}$), the $k^{-4}$-scaling marks the energy dissipation within vortices, which is dominated by viscosity~\cite{Alert2021review}. Meanwhile, $k^{-2}$ and $k^{-1}$ represent two mutually exclusive scaling laws, which govern the transitional regimes in the $k$-space for 2D and 3D samples respectively. The $k^{-2}$-scaling governs the regime $D_v^{-1}\ll k \ll H^{-1}$ and becomes evident in the 2D limit~($H^{-1}\gg k, D_v^{-1}$). On the other hand, the $k^{-1}$-scaling governs $H^{-1}\ll k \ll D_v^{-1}$ and emerges in the 3D limit ($H^{-1}\ll k, D_v^{-1}$). In all, our model supports the existence of universal scaling for active turbulence in both the 2D and 3D limits. 

Consistently, our experimental results agree with the predicted universal scaling. In 3D samples, the measured scaling $(\gamma_1,\ \gamma_2)=(-1, -4)$ (\figref{fig:TKE}c), and $k^1$-scaling are observed at small wavenumbers (\extfigref{fig:TKELowK}). These all align perfectly with the prediction, see the right vertical dashed line in \figref{fig:theory}c. For samples in the 2D-3D crossover regime, e.g., around $H=H_1$, the kinetic energy spectrum is predicted to evolve from $k^1$-scaling to $k^{-4}$-scaling with no transitional scaling (middle vertical line in \figref{fig:theory}c). It is exactly what we observe $(\gamma_1,\ \gamma_2)=(1, -4)$ around $H=H_1$, see \figref{fig:TKE}b and e. Lastly, in the 2D limit, only the $k^1$- and $k^{-2}$-scaling are observed (\figref{fig:TKE}a) because the predicted $k^{-4}$-scaling only exists on the scales smaller than the bacterial size. On such scales, the bacterial suspensions cannot be regarded as a continuum and thus the model becomes not applicable, as is shown by the gray shaded area in \figref{fig:theory}c. 

\section*{Discussion and summary}
In summary, we have observed the universal scaling behaviors of bacterial turbulence in the 2D and 3D limits, which are independent of bacterial activity, bacterial length, and system height. The universal scaling laws are confirmed theoretically by considering the hydrodynamics of active fluids confined between two no-slip walls. From 2D to 3D, the active turbulence exhibits a two-step transition: spectral scaling governing the small length scales changes at $H_0$ and a characteristic $k^{-1}$-scaling for 3D turbulence emerges after $H_1$. The two critical heights are also found in evolution of real-space structures. $H_0$ emerges from the interplay of the confinement height $H$ and the cell length; while $H_1$ reflects the competition between the system height and the vortex size. \rev{Lastly, a characteristic $H^{0.5}$-dependence of vortex size is uncovered experimentally for 3D bacterial  turbulence and is derived using stability analysis in confined active polar systems.}

Our hydrodynamic model considers active fluids as a continuum force field and use image systems to capture the hydrodynamic screening effect of the no-slip boundaries as the leading-order approximation. Here we do not consider the non-uniformity in the density distribution, which could be important in presence of giant density fluctuations~\cite{Narayan2007, Zhang2010} and motility-induced phase separation (MIPS)~\cite{Cates2015}. Also, the inner structure of the vortex is largely simplified in the sense that the continuum assumption ignores irregular shapes and complex kinetics of realistic vortices at small scales (high-$k$). Nevertheless, the model still captures our experimental findings excellently, showing that the details above have only minor effects in determining the scaling properties of active turbulence. One possible reason is that bacterial turbulence is dominated by hydrodynamic interaction~\cite{Peng2021}, which suppresses the density fluctuation and MIPS~\cite{Cates2015}. 
In addition, our experiments confirm that the exponential vortex size distribution, which applies for active nematic turbulence~\cite{Giomi2015,Martinez-Prat2021prx}, is also a good approximation in bacterial turbulence. Hence, we believe our theoretical model captures the essential elements of scaling behaviors of active turbulence from 2D to 3D. The model can be applied in hydrodynamic-dominated active systems where the vortex-vortex interaction is negligible and MIPS is absent.

The scaling laws found in our experiments and theories show wide applicability. The energy spectra and vortex structures reported by other experiments that employ different bacteria species and/or are conducted at higher concentrations~\cite{Wensink2012,Liu2020} are consistent with both our experimental results and theoretical predictions. Even in dense suspensions of mammalian sperms where the active agents are an order of magnitude larger than bacteria and are of a distinct propulsion mechanism, the energy spectrum still displays $k^{-4}$-scaling when $k \gg 1/H, 1/D_v$, in line with our results~\cite{Creppy2015spermTurbulence}. Lastly, it is noteworthy that the existence of the four scaling regimes does not rely on a specific $D_v-H$ relation but only requires that $D_v$ increases with $H$ slower than linearly in the 3D limit (so that $1/H$ and $1/D_v(H)$ can intersect). As the $H^{0.5}$ scaling is also reported in active nematic turbulence~\cite{Brotto2013, Wioland2016, Chandrakar2020prl, Varghese2020}, we expect these systems to exhibit similar four scaling regimes from 2D to 3D.

Besides hydrodynamic-interaction-dominated bacterial turbulence confined between two no-slip solid surfaces, the collective motion of active matter systems can be dominated by other interactions, like short-ranged steric interaction~\cite{Patteson2018} and long-ranged electromagnetic interactions~\cite{Meng2018,Meng2021prl}. The confining boundary could be liquid-liquid interfaces~\cite{Martinez-Prat2021prx} or liquid-air interfaces~\cite{Peng2016}, which provide distinct couplings within active fluids. How other interactions and boundary types affect the universal scaling is a question worth further exploring.

\section*{Acknowledgments} 
Y.P. acknowledges the support from the National Natural Science Foundation of China (NSFC) (No. 12074406 and No. T2221001). F.M. acknowledges supports from the National Natural Science Foundation of China (NSFC) (No. 12047503 and No. 12275332), Chinese Academy of Sciences (No. XDPB15), Max Planck Society (Max Planck Partner Group) and Wenzhou Institute (WIUCASICTP2022). 

\section*{Author Contributions}
Y.P. and F.M. conceived the project. D.W. performed the experiments and analyzed the data. F.M., R.G., Y.Y. and X.W. conducted the theoretical modeling and calculations. D.W., Y.P. and F.M. wrote the paper with input from M.L. and R.G.. All authors discussed the results and reviewed the manuscript. 

\appendix

\section{Methods} 
\subsection{Fluidic chamber}
\begin{figure}[ht!]
    \includegraphics[width=0.48\textwidth]{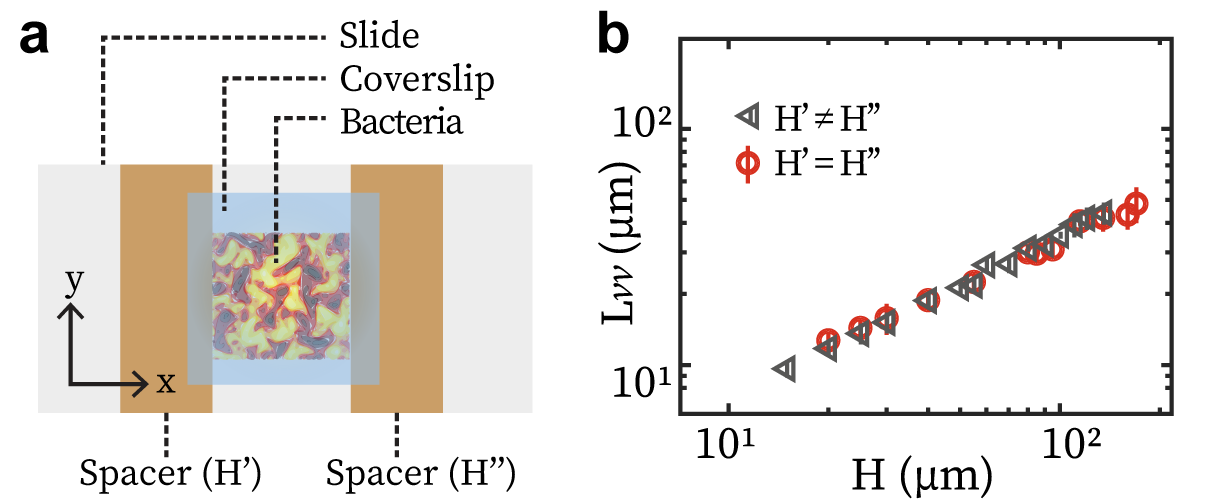}
    \vspace{-2mm}
    \caption{\textbf{Experimental samples.}(a) Schematics of a chamber. (b) Velocity correlation length measured in the sloped chambers ($H'\neq H''$) and uniform-height chambers($H'=H''$). Symbols and error bars represent mean and one standard deviation respectively.}
    \label{fig:chamber}
\end{figure}

We make fluidic chambers with glass slides and coverslips as shown in \extfigref{fig:chamber}a. After being filled with bacteria suspension, chambers are sealed with UV-cured glue (NOA63, Norland Optics Inc.). To tune the chamber heights, we use polystyrene beads as spacers for chamber height $H<10$~\um, and different tapes for chambers of $H\gtrsim7$~\um. Besides chambers of uniform height, sloped chambers are made using tapes of different heights ($H' \neq H''$) and are calibrated per piece. They allow fine tune over $H$ and high-throughput experimentation. In such chambers, the bottom and top plates are approximately parallel (incline angle $<0.4\degree$). Turbulent statistics measured in those two types of chambers are same, see \extfigref{fig:chamber}b.

\clearpage
\subsection{Bacterial suspension}
We employ a wild-type \textit{E.coli} strain (BW25113) in our experiment. To extend the swimming period in closed chambers, we insert a light-driven transmembrane proton pump, proteorhodopsin (PR), into the strain~\cite{Walter2007PRecoli}. A typical cell body is a 3~\um-long rod with 0.8~\um in diameter. We first culture bacteria in Terrific Broth (TB) culture medium overnight (12-16~h) at 37~$\degree$C. The saturated suspension is diluted for 100 times with fresh TB medium and then is further cultured at 30~$\degree$C for 6-8~h. We harvest bacteria and centrifuge them at $800-1000$~g for 5 min. The bacteria are resuspended in adjusted Berg's motility buffer, containing 6.2~mM K$_2$HPO$_4$, 3.8~mM KH$_2$PO$_4$, 67~mM NaCl, 0.1~mM EDTA, 4.5$\times10^{-3}$\%~(v/v) TWEEN20, and 0.5\%~(w/v) glucose. The final body volume of bacteria is $\phi\approx0.06$ (number density $n\approx3.2\times10^{10}$~cells/ml). 

\subsection{Cephalexin treatment}
\begin{figure}[htbp!]
    \includegraphics[width=0.48\textwidth]{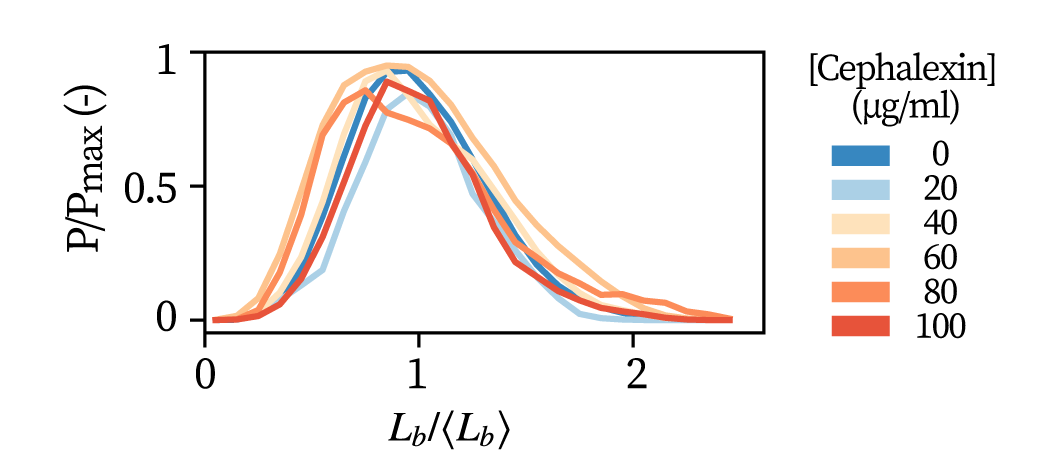}
    \vspace{-2mm}
    \caption{\textbf{Bacterial length distributions after cephalexin treatment.} Distribution of scaled cell length $L_b/ \langle L_b \rangle$. Distributions are normalized by their respective maxima.}    
    \label{fig:cellLengthHistogram}
\end{figure}

Bacterial length is tuned by inducing cephalexin (Solarbio Inc.) of different concentrations ($\rm{[Cephalexin]}$) at 3.5~h after the second stage of culturing begins. We measure the length for N$\approx$1000 cell at each $\rm{[Cephalexin]}$. The length polydispersity is unvaried, see \extfigref{fig:cellLengthHistogram}. Additionally, bacterial motility is not affected by the cephalexin treatment. The mean swimming speed of single cells are measured in H=8~\um samples (N$\approx$20 for each concentration) and they remain constant in the tested antibiotic concentrations, in line Ref.~\cite{Nishiguchi2017cephalexin,Li2019bacteriaNematics}. In the series of experiments involving cephalexin, we maintain the cell body volume fraction $\phi\approx0.06$. 

\subsection{Flow field extraction}
Bright field microscopy are performed with an inverted microscope (Nikon Ti2e). A $40\times$ objective is employed primarily for measurement and a $10\times$ objective is used to obtain a large field of view, see \extfigref{fig:TKELowK}. Videos are recorded with an sCMOS camera at 30~fps (PCO.edge). A typical measurement lasts for 5~s. For each sample of uniform height, $\sim$20 measurements are taken; while in the sloped chamber, $\sim$8 measurements are performed at each height (location). Recorded videos are analyzed with particle image velocimetry (\textsc{MATLAB} PIVLab2.37~\cite{Thielicke2014PIV}). We use windows of $16\times16$ pixels with 50\% overlap for convolution. 

\subsection{Measuring correlation length for \textbf{$H=$}400 \um}
For the thickest samples measured ($H=400~$\um, empty red marker at the right extreme of \figref{fig:Lvv}a), flow patterns at $d_c=H/2$ are too obscured to extract. The presented values are asymptotic estimations by extrapolating the trend measured with $\rho=20n_0$ from $d=0-120$~\um, and $\rho=40n_0$ from $d=0-70$~\um.

\begin{figure}[htbp!]
    \includegraphics[width=0.48\textwidth]{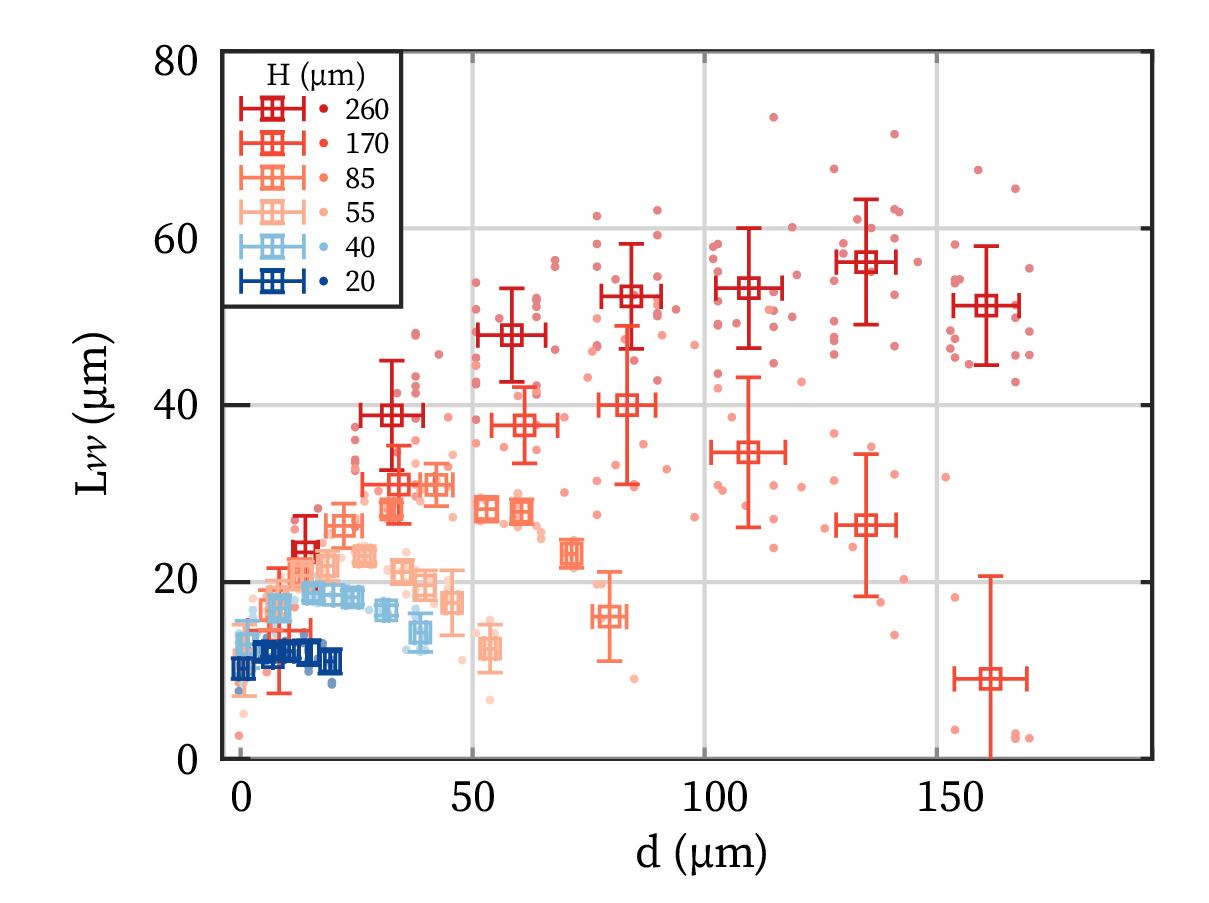}
    \vspace{-2mm}
    \caption{\textbf{Variation of the correlation length in $z$.} Measurements of $L_{vv}$ as a function of $d$ (N=3-5 samples) are plotted as dots. Symbols with error bars represent the mean and standard deviation of the binned data.}
    \label{fig:LvvOverD}
\end{figure}

\subsection{Data pooling and averaging}
\begin{figure}[htbp!]
    \includegraphics[width=0.48\textwidth]{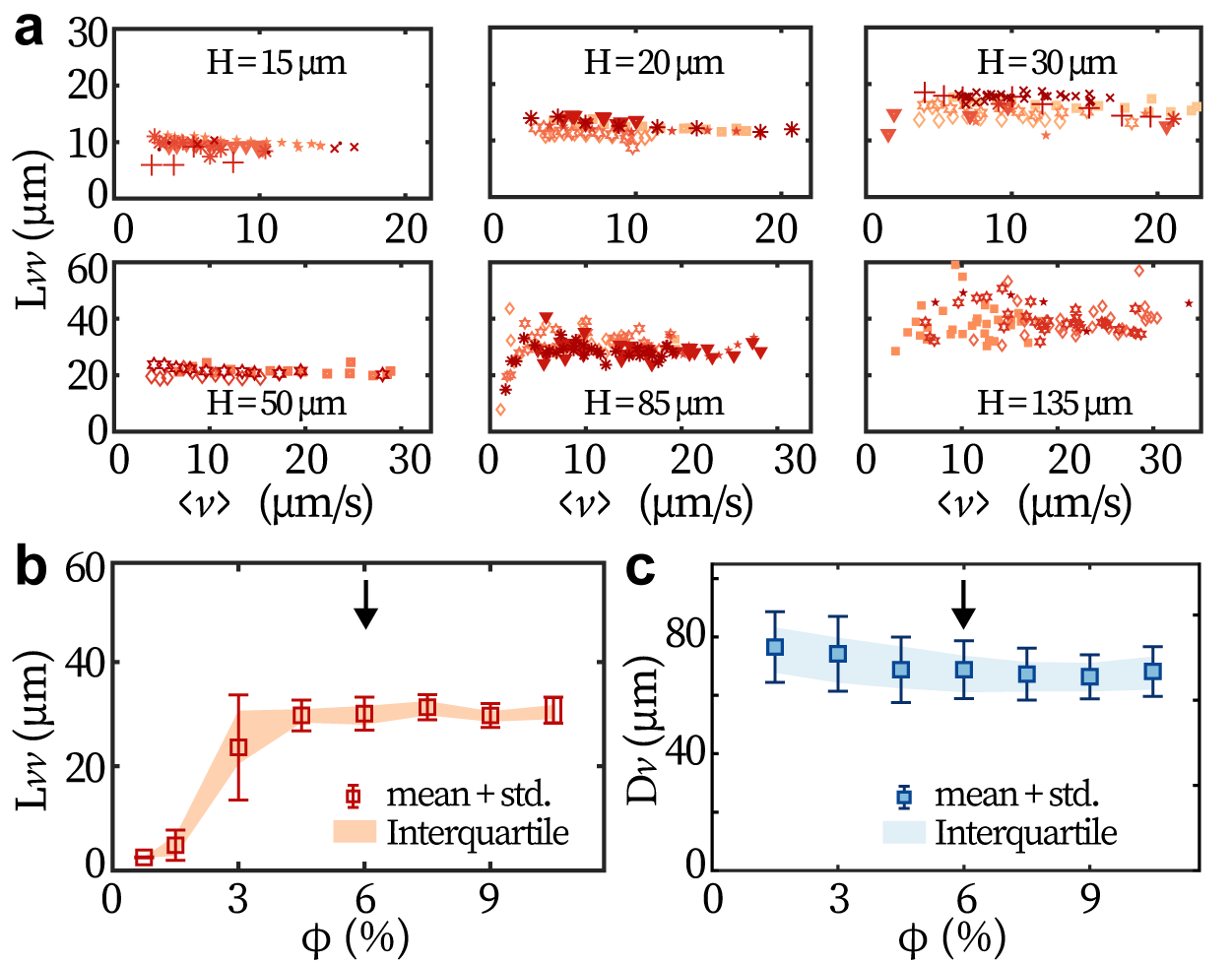}
    \vspace{-5mm}
    \caption{\rev{\textbf{Independence of vortex structure on activity and bacterial density } (a) Measurements of $L_{vv}$ at different bacterial activity $\langle v \rangle$. Different symbols represent different experiments. Measurements are perform at $d_c=H/2$. (b) $L_{vv}$ as a function of bacterial volume fraction $\phi$ ($H$=85\um, taken at $d_c=H/2$). (c) $D_{v}$ as a function of $\phi$. Black arrows mark the bacterial density used for the presented data in the main text.}}
    \label{fig:activityIndependence}
\end{figure}

For each confinement height $H$, the highest possible mean turbulence speed $\langle v \rangle_{\rm max}$ is employed to benchmark the activity level of a particular bacterial suspension, see \figref{fig:TKE}a-c. $\langle v \rangle_{\rm max}$ is obtained from measurements over multiple samples, which are prepared following the same protocol. The energy spectrum of a particular $H$ displayed in \figref{fig:TKE}d-f and \figref{fig:cellLength} is averaged over $\langle v \rangle/ \langle v \rangle_{\rm max}>0.5$. Although $L_{vv}$ and $D_v$ are found to depend negligibly on $\langle v \rangle$ after the onset of turbulence \rev{(\figref{fig:activityIndependence}a)}, same as reported previously~\cite{Sokolov2012,Dunkel2013,Heidenreich2016pre}.  However, for consistency, the same criterion on activity is applied for $L_{vv}$ and $D_{v}$. In measuring $L_{vv}$ as a function of $d$, we set no criteria for bacterial activity as $\langle v \rangle$ varies with $d$. The raw measurements of $L_{vv}(d)$ are displayed in \figref{fig:LvvOverD}.

\section{Independence of turbulence structure on bacterial density}
\rev{Bacteria are known to accumulate near no-slip boundaries~\cite{Berke2008}. Such accumulation may induce number density variation along the system's $z$ axis. If the characteristic lengths of turbulence (e.g., $L_{vv}$ and $D_{v}$) depend on bacterial number density, then the near-wall accumulation will induce uncertainties in the reported data, for example, in $L_{vv}$'s dependence on $d$ (\figref{fig:Lvv}). However, we show that bacterial number density matters negligibly when it is sufficient for turbulence to emerge. Both $L_{vv}$ and $D_{v}$ remain almost constant after the bacterial volume ratio $\phi$ exceeds $\sim$3\%, see \figref{fig:activityIndependence}b and c respectively. The presented data are measured at $d_c=H/2$, and we confirm that the trends are qualitatively the same for other heights ($d$).}

\section{Critical heights measured in different representations.}
\begin{figure}[htbp!]
    \includegraphics[width=0.45\textwidth]{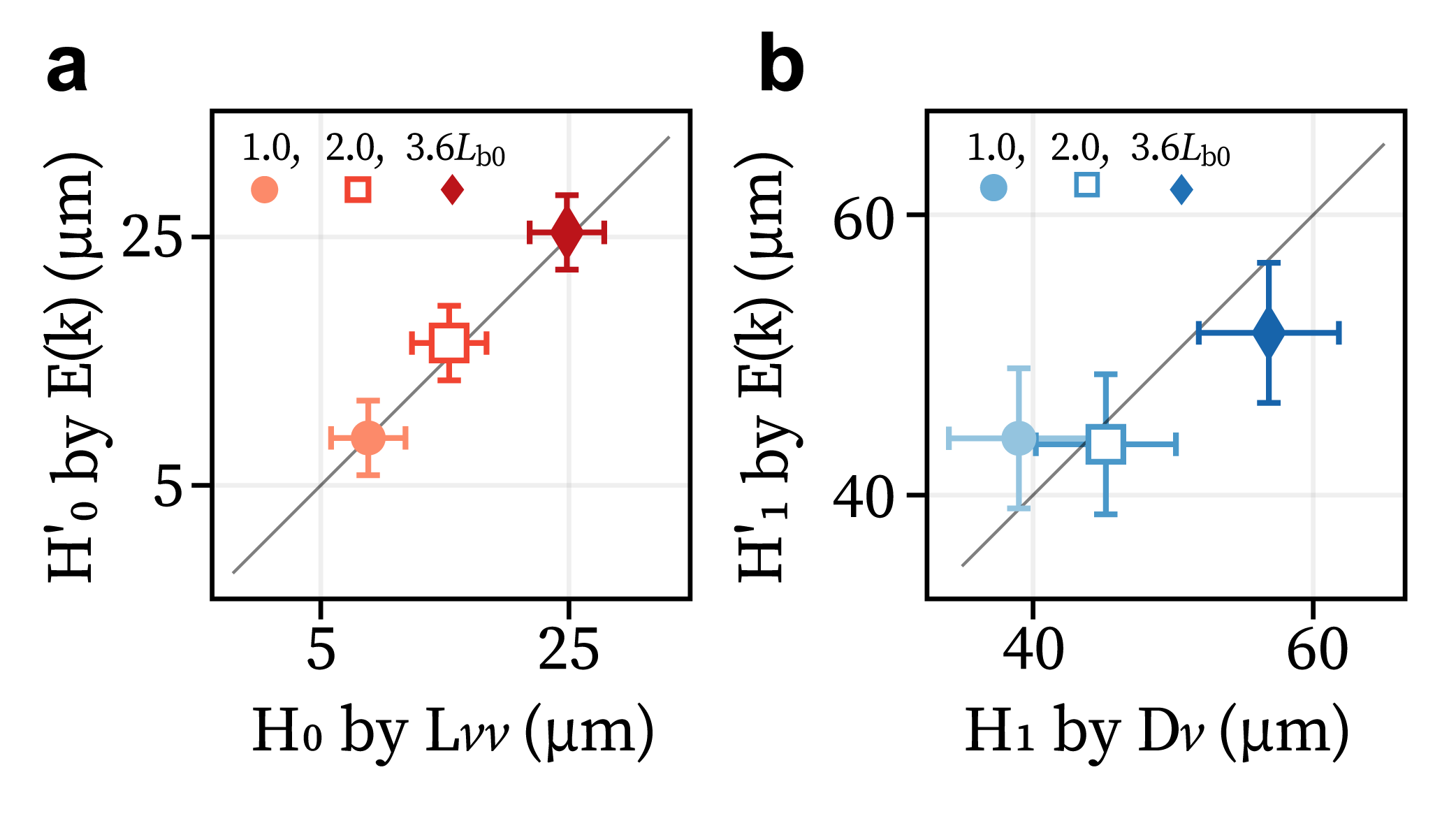}
    \vspace{-2mm}
    \caption{\textbf{Critical heights extracted from velocity correlation functions and from energy spectra}. $H_0$ (a) and $H_1$ (b) measured in different representations. The solid lines are the identity line ($y=x$). The error bars in the $x$ and $y$ axes represent the uncertainties in $H$.}
    \label{fig:H0H1equivalence}
\end{figure}

The critical heights $H_0$ and $H_1$ manifest both in the characteristic lengths of the turbulence against $H$ ($L_{vv}$ and $D_v$) and in the kinetic energy spectra. Here we show that the measured values match to each other. $H_0$ and $H_1$ measured in different representations are displayed in \extfigref{fig:H0H1equivalence}a and b respectively. The $x$-axis of \extfigref{fig:H0H1equivalence}a shows $H_0$ measured as where $\gamma_2$ drops below -2 in the energy spectra; while the $y$-axis is $H_0$ measured as the $x$-intercept of the linear fitting to the initial fast-increasing trend in $L_{vv}$. The $x$-axis of \extfigref{fig:H0H1equivalence}b reports $H_1$ measured in energy spectra as where $\gamma_1$ drops below 1. The $y$-axis reports $H_1$ as where $H=D_v$ (with linear interpolation). Data gather around the identity lines ($y=x$, gray lines), showing that the two approaches are equivalent.

\begin{figure}[htbp!]
    \includegraphics[width=0.47\textwidth]{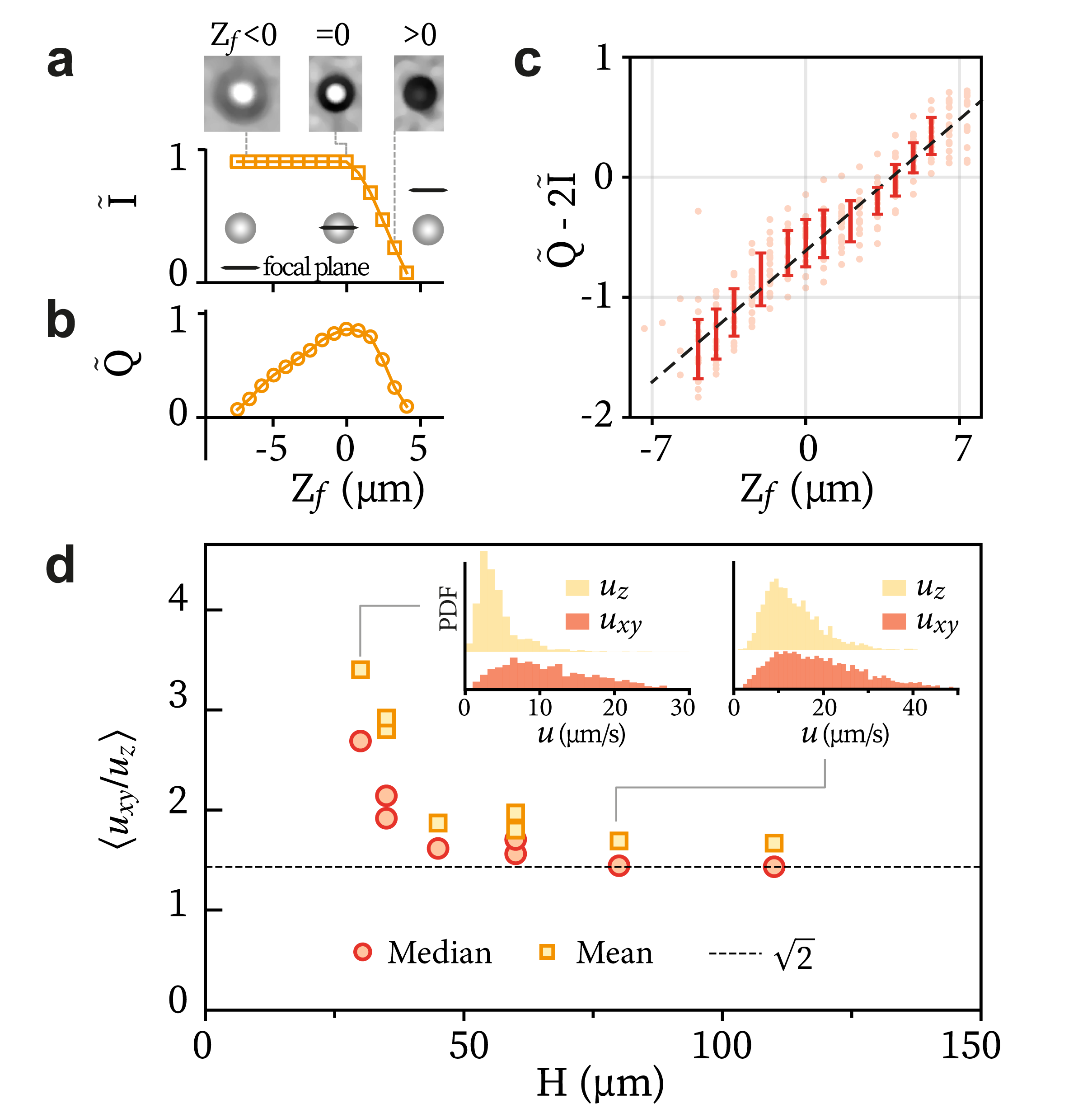}
    \vspace{-2mm}
    \caption{\textbf{Enhanced particle tracking velocimetry to measure 3D flows.} Scheme of extracting the $z$-position of beads in particle tracking velocimetry (PTV). (a) The normalized intensity ($\tilde{I}$) and (b) tracking quality factor ($\tilde{Q}$) of a fixed bead. Upper images display the beads at different relative heights $Z_{f}$. Insets: relative heights between the focal plane (solid line) and the bead, corresponding to the upper images respectively. (c) The composite metric $\tilde{Q}-2\tilde{I}$ corresponds linearly to the $z$-position. Dots: single beads; error bar: mean$\pm$std; dashed line: linear fit. (d) The mean ratio of the $z$-component of speed, $u_{z}$, to the $xy$-component of speed, $u_{xy}$. Insets: representative histograms of $u_{z}$ and $u_{xy}$ in samples of $H=30$~\um (left) and 80~\um (right). Histograms are vertically shifted for clarity.}
    \label{fig:PTV}
\end{figure}

\section{Measuring vertical flow using enhanced particle tracking velocimetry.}
To confirm the velocity field has become isotropic at $H=H_1$, we measure the $z$-component of speed $u_z$ in the following way. We first develop an enhanced particle tracking velocimetry (PTV) that extracts a bead's $z$-axis position by its appearance. The scheme of the technique is displayed in \extfigref{fig:PTV}a-c. Beads are tracked with the TrackMate plugin of ImageJ and the algorithm primarily depends on fitting the beads' interior with a Laplacian-of-Gaussian (LoG) kernel. We exploit the mean pixel intensity $I$ of the beads' interior and the quality factor $Q$ which measures how well the LoG kernel fits. \extfigref{fig:PTV}a-b display how the normalized metrics ($\tilde{I}$ and $\tilde{Q}$) varies with the focal plane height ($Z_f$, defined as shown in the schematic drawing). $\tilde{I}$ saturates at $\sim$1 when $Z_f\leq0$ and drops for $Z_f>0$, i.e. the bead center turning from white to black. On the other hand, $\tilde{Q}$ peaks at $Z_f=0$, i.e. the bead that is in-focus has the highest quality factor. These two trends are illustrated by the images of beads at the top of \extfigref{fig:PTV}a. Combine these two metrics as $\tilde{Q}-2\tilde{I}$, we obtain a metric that translates linearly to $Z_f$, see \extfigref{fig:PTV}c.

Practically, we employ polystyrene beads of $a=6$~\um diameter for PTV. The relationship between $\tilde{Q}-2\tilde{I}$ and $Z_f$ is calibrated by $\sim50$ beads per sample after the bacterial suspension has ceased moving. Such configuration allows us to track all three coordinates of the beads in a slice of $\sim2a=12$~\um thick. When a beads goes out of this slice, a track is considered finished. We pool $2\times10^3-1\times10^4$ tracks per sample. All tracks are broken down as 1 s long segments. This duration for each segment is determined by the characteristic time scale of the temporal velocity auto-correlation (1-3 s). The statistics of the speed (total arc length/1 s) in the $z$-axis and that in the $xy$-plane are displayed in the insets of \extfigref{fig:PTV}d. Altogether, the mean or median of $u_z/u_{xy}$ converges to $\sqrt{2}$ when $H=40-50$~\um, which matches with $H_1$ observed from the vortex size (\figref{fig:Lvv}c) or kinetic energy spectra (\figref{fig:TKE}h). The ratio supports that bacterial turbulence turns isotropic ($\langle u_x \rangle\approx\langle u_y \rangle\approx\langle u_z \rangle$).

\section{Energy spectrum at low-$k$}

\begin{figure}[htbp!]
    \includegraphics[width=0.48\textwidth]{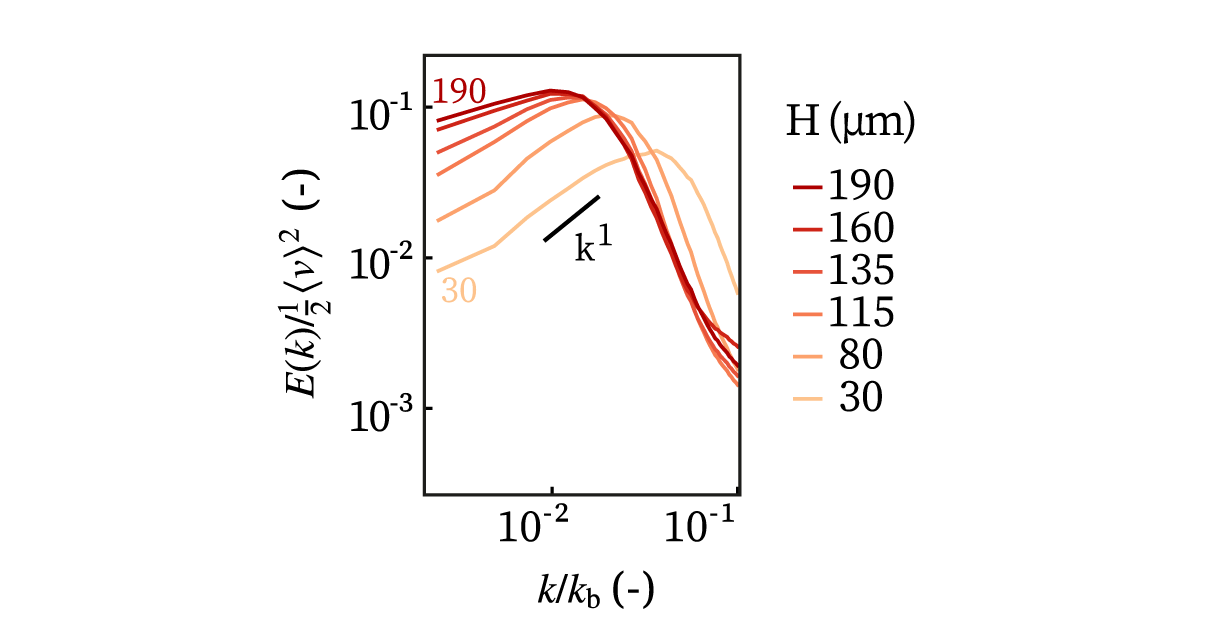}
    \vspace{-5mm}
    \caption{\textbf{Kinetic energy spectrum at the low-$k$ end.} Flow fields acquired with lower microscope magnification (10$\times$) are used to probe the spectral scaling at the low wavenumbers.}
    \label{fig:TKELowK}
\end{figure}

Our theory predicts that the scaling $E(k) \sim k$ dominates the low-$k$ end of the spectrum. To observe examine such spectral behavior for $H>50$ \um, we measure samples with 10$\times$ magnification. Under this low magnification, we are able to extract the flow field in samples up to $H$=190 \um, see \figref{fig:TKELowK}. The $E(k)\sim k$ scaling regime presents clearly while the spectra continue shifting to the left (\figref{fig:TKE}d-e).

\section{Exponential distribution of vortex size}

\begin{figure}[htbp!]
    \includegraphics[width=0.45\textwidth]{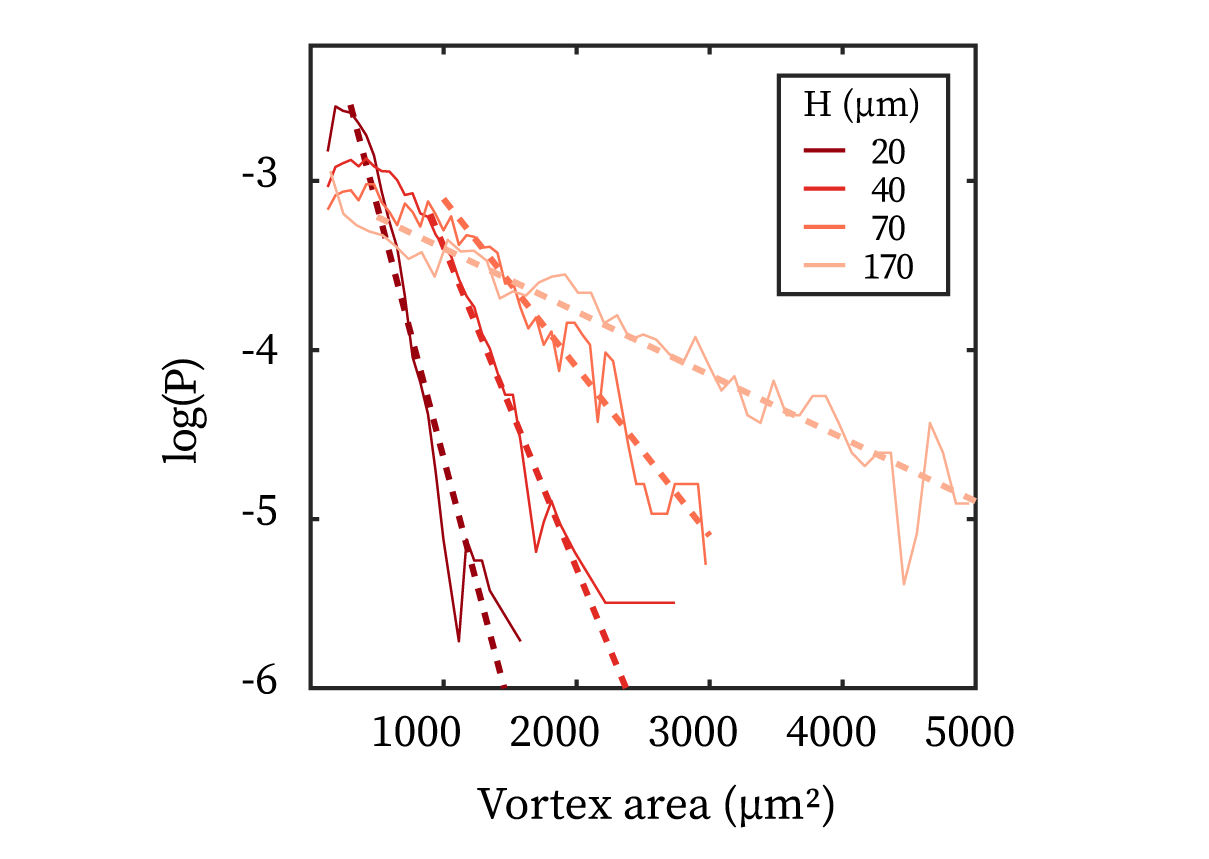}
    \vspace{-5mm}
    \caption{\textbf{Exponential distribution of vortex size.} Vortex size distribution in samples of different heights. Cross-sectional sizes of the vortices in the $xy$-plane are extracted with the algorithm described in Ref.~\cite{Giomi2015}. Dashed lines are the exponential fits.}
    \label{fig:ExpDistribution}
\end{figure}

To validate one of theoretical assumptions that vortices in bacterial turbulence follow exponential size distributions, we measure vortex sizes with the algorithm used in Ref.\cite{Giomi2015}. The algorithm first computes the Okubo-Weiss field $Q=-\partial_x u_x\cdot\partial_y u_y + \partial_x u_y \cdot \partial_y u_x$ from numerical data of the flow field. A region with $Q<0$ and containing a flow circulation (vortex core) is considered as a vortex. Areas of such regions measure the vortex size projected in the $xy$-plane. We implement this algorithm in samples of varying $H$ and confirm that the vortex size follows exponential distributions, see \figref{fig:ExpDistribution}.

\clearpage
 \bibliographystyle{naturemag}
 \bibliography{reference}

\end{document}